\documentclass[epj]{svjour}

\usepackage{latexsym}
\usepackage{graphics}
\usepackage{hyperref}
\usepackage{indentfirst}
\usepackage{bm}
\usepackage{cite} 
 \hypersetup{colorlinks=true, urlcolor=blue, 
linkcolor=blue, citecolor=blue} 
\usepackage{lineno,hyperref}
\usepackage{graphicx}
\usepackage{amsmath}
\usepackage{amssymb}
\usepackage{mathrsfs}
\usepackage{tikz}
\usepackage{filecontents}
\usepackage{siunitx}

\begin{document}
\title{Enhancing Electrical Properties and Seebeck Effects of WO$_3$ Thin Films Using Spray Pyrolysis: Insights into the Conductivity and Carrier Type}
\author{
Zahra Asghari \inst{1} 
\thanks{\emph{}
e-mail: zahraasghary42@yahoo.com}
 \and Hamid Arian Zad \inst{2}
 \thanks{\emph{}
e-mail: arianzad.hamid@yerphi.am}
  \and Hosein Eshghi \inst{1}
}

%
%
\institute{Physics Department, Shahrood University of Technology, Shahrood, Iran
\and Alikhanyan National Science Laboratory, Alikhanian Br. 2, 0036 Yerevan, Armenia 
       }

%
%
\abstract{
This study focuses on enhancing the electrical and thermoelectric properties of tungsten trioxide (WO$_3$) thin films using the spray pyrolysis technique. Three samples, namely A$_1$, A$_2$, and A$_3$, were prepared by depositing the films onto glass substrates at different deposition rates: R$_1=1$ mL/min, R$_2=3$ mL/min, and R$_3=7$ mL/min, respectively.
Subsequently, the films underwent annealing at 500 $^\circ$C for 2 hours in air. A detailed investigation was conducted to analyze the structural, morphological, optical, and electrical properties of the $\mathrm{WO}_3$ thin films.
The FE-SEM images revealed the formation of circular strings with varying diameters for all layers. The diameter of each string decreased as the deposition rate increased. XRD structural analysis indicated that, before annealing, sample R$_1$ exhibited a polycrystalline nature with mixed tetragonal-hexagonal phases, while the other two samples were amorphous. After annealing, sample A$_3$ also became polycrystalline. Moreover, the UV-Vis spectra analysis demonstrated a monotonic decrease in the band gap of the layers with increasing spray rate, potentially attributed to the attenuation of the quantum confinement effect.
The Seebeck effect was used to authenticate the $n$-type conductivity of the WO$_3$ layers, as the Seebeck effect and conductivity exhibited a direct connection. Furthermore, the current-voltage variation was found to align with the oxygen vacancies sites. Remarkably, the sample A$_3$ with the highest deposition rate exhibited the lowest resistivity.
 }

\PACS{
      {}{}   \and
\\
\textbf{Keywords: }{Tungsten trioxide WO$_3$; Seebeck effect; Thermoelectrical properties}
     } 
%

\maketitle

\section{Introduction} \label{sec:level1}

Semiconductor characterization plays a crucial role in the advancement of semiconductor technology. Among the intermediate metal oxides, tungsten trioxide ($\mathrm{WO}_3$) stands out as a widely studied $n$-type semiconductor with a large indirect band gap of approximately 2.4-3.7 eV \cite{Dalavi2011,Zhang2011,Kim2009,Rao2013,Marques2018,Tian2017}. Its optoelectronic properties have attracted considerable attention for potential applications such as smart windows \cite{Marques2018,Tian2017,Mouratis2022}, front contacts in photovoltaic solar cells \cite{Razmi2011,Gullapalli2010,Ramos2007}, and variable-reflectance mirrors \cite{Zan2019,Kalhori2017}.

Thin films of $\mathrm{WO}_3$ have recently gained significant prominence due to their excellent electrochemical stability, making them ideal for use as photo-anodes in photo-electrochemical cells \cite{Hu2017,Wang2019,Chang2021_1,Wang2017_1,Zheng2018_1}. There are  other works that provide insights into the fabrication techniques, properties, performance enhancement strategies, and potential applications of $\mathrm{WO}_3$ thin films in the field of gas sensing \cite{Sivathas2022,Lu2018}. Previous studies have reported the intrinsic reaction mechanism of $\mathrm{WO}_3$/electrolyte interfaces in various electrolytes \cite{Heieh2010,Cole2008,Granqvist2000,Sella1998}. The deposition of $\mathrm{WO}_3$ thin films has been achieved through various techniques, including physical vapor deposition (PVD), chemical vapor deposition (CVD) \cite{Bertus2012}, evaporation \cite{Kovendhan2011,Lozzi2001,Cantalini2000}, electrodeposition \cite{Porqueras2000}, sputtering \cite{Livahe1996,Ronnow1996,Guillen2023,Guillen2023_AECM}, sol-gel deposition \cite{Cantalini2000,Wang1999}, and spray pyrolysis technique \cite{Regragui2000,Regragui2003,Kamberov2019,Mohite2015}. The latter technique presents many advantages such as: (i) short time deposition, (ii) simple technology, (iii) low-cost, (iv) no need for vacuum, and (v) large scale area of thin film production.

Many researchers have devoted themselves to studying $\mathrm{WO}_3$ thin films prepared under different spray rates \cite{Porqueras2000} and the effect of annealing on the structural, surface morphological, and optical properties of $\mathrm{WO}_3$ thin films \cite{Kovendhan2011,Livahe1996,Wang1999,Abe2008,Qamar2010}. For example, the $p$-type conductivity of $\mathrm{WO}_3$ thin films has been considerably investigated \cite{Bertus2012}. The materials were deposited by spray pyrolysis using a precursor tungsten hexachloride dissolved in ethanol. The authors of the same reference sagely considered parameters such as deposition temperature and pressure of the carrier gas as typical variables, resulting in enhanced optical and electrical properties of the $\mathrm{WO}_3$ thin films. The influences of deposition parameters on surface morphology, conductivity, and optical and electrochromic properties were studied using X-ray diffraction (XRD), current-voltage measurements ($I-V$), UV-Vis spectrometry, and other techniques \cite{Bertus2012}.

The Seebeck effect, also known as the thermoelectric effect, is of great interest in semiconductor physics, as it allows for the determination of the type of dominating charge carriers in
 a material and the relative position of the Fermi level with respect to the transport level. Phenomenologically, the Seebeck effect can be understood as a temperature gradient-induced voltage generation due to the shift in the energetic distribution of free charge carriers in a metal or semiconductor. However, the investigation of the Seebeck effect in $\mathrm{WO}_3$ has been relatively limited.

In this paper, we present a detailed investigation of the Seebeck effect in $\mathrm{WO}_3$ thin films. Our study focuses on three different samples, examining the thermoelectrical properties of $\mathrm{WO}_3$ and shedding light on the anomalous behavior of this fascinating material under different deposition rates and annealing conditions. We verify the influence of the deposition rate and annealing process on various physical properties of the $\mathrm{WO}_3$ thin films, including conductivity, optical properties, and electrical behavior. Furthermore, we analyze the correlation between variations in the Seebeck effect and changes in the aforementioned properties across the different samples. Through this comprehensive analysis, we aim to gain a deeper understanding of the thermoelectrical properties of $\mathrm{WO}_3$ and provide valuable insights for its potential applications \cite{Porqueras2000,Kovendhan2011,Livahe1996,Wang1999,Abe2008,Qamar2010,Bertus2012}.
\section{Experimental Procedure: General Information}

The deposition of $\mathrm{WO}_3$ thin films on glass substrates was carried out using the spray pyrolysis method. The precursor solution consisted of 50 mL of ammonium tungstate [$(\mathrm{NH}_4)_2\mathrm{WO}_4$], prepared by dissolving pure $\mathrm{WO}_3$ powder (99.9$\%$, Merck) in a mixture of ammonia and distilled water. Three different spray rates, denoted as R$_1=1$ mL/min, R$_2=3$ mL/min, and R$_3=7$ mL/min, were used for the deposition. The other deposition parameters, including the spray solution volume, nozzle-to-substrate distance, and substrate temperature, were maintained at 50 mL, 30 cm, and 400 $^\circ$C, respectively.

Prior to deposition, all substrates underwent a cleaning procedure. They were washed with soap and water, followed by immersion in a container containing acetone and distilled water. Subsequently, the substrates were placed in an ultrasonic machine for 16 minutes. Finally, the substrates were rinsed with deionized water and dried using nitrogen gas.

The preparation of transparent $\mathrm{WO}_3$ thin films involved two steps. In the first step, a solution of ammonium tungstate [$(\mathrm{NH}_4)_2\mathrm{WO}_4$] was sprayed onto the preheated substrates (at 400$^\circ$C) using air as a gas conveyor. The experimental setup for the spray deposition process has been previously described in Ref. \cite{Perednis2003,Perednis2005,Jeyadheepan2019}. The resulting films exhibited transparency, amorphous structure, homogeneity, and good adhesion to the substrate. In the second step, the sprayed films were annealed in ambient air. The annealing temperature was fixed at 500 $^\circ$C, and the annealing time was set to 2 hours. The annealed films appeared transparent with a slightly yellowish color. The deposition of the $\mathrm{WO}_3$ thin layer followed the chemical reactions \cite{Sivakumar2004}:
\begin{equation}\label{WO3_CR}
\begin{array}{lcl}
\mathrm{WO}_3+2\mathrm{NH}_3+H_2\mathrm{O} \xrightarrow{400^\circ \text{C}} (\mathrm{NH}_4)_2\mathrm{WO}_4, \\
(\mathrm{NH}_4)_2\mathrm{WO}_4 \xrightarrow{400^\circ \text{C}} \mathrm{WO}_3+\mathrm{H}_2\mathrm{O}+2\mathrm{NH}_3.
\end{array}
\end{equation}

The morphologies of the samples were investigated using a Field Emission Scanning Electron Microscope (FE-SEM) (SIGMA 300-HV, Carl Zeiss). The structural characterization of the films was performed using X-ray diffraction (XRD) with CuK$_\alpha$ radiation (wavelength of 1.54 $\si{\angstrom}$) and an angular range of 2$\theta$ from $10$ to $70$ degrees (XRD; Bruker AXS). The transmittance spectra of the studied samples were measured using a UV-Vis spectrophotometer (Shimadzu UV-Vis 1800) in the wavelength range of $300-1100$ nm. The film thickness was determined using a Taylor/Hobson profile meter with an accuracy of +20 nm.

\section{Results and discussion}
The Seebeck effect, also referred to as the thermoelectric effect, is a phenomenon that occurs when there is a temperature difference across a conductor, such as a metal or semiconductor, resulting in the generation of an electric voltage. This effect is based on the behavior of charge carriers, such as electrons or holes, in response to temperature gradients.
When a temperature gradient is applied to a conductor, meaning one end is heated while the other end is cooled, the charge carriers experience an imbalance in their distribution. Due to their thermal energy, the charge carriers diffuse from the hotter region to the colder region, seeking equilibrium. As a result, an accumulation of charge occurs at both ends of the conductor, creating an electric field.
The electric field gives rise to a potential difference or voltage between the two ends of the conductor, which can be measured as the Seebeck voltage. This voltage is directly proportional to the temperature difference across the conductor. Hence, a larger temperature gradient will result in a higher Seebeck voltage.
The Seebeck effect is fundamental to thermoelectric devices, which utilize this phenomenon to convert heat energy directly into electrical energy or vice versa. These devices, known as thermoelectric generators or thermoelectric coolers, find applications in various fields, including power generation, waste heat recovery, and temperature control systems.

Overall, the Seebeck effect demonstrates the conversion of a thermal gradient into an electric potential, highlighting the interconnected nature of temperature and electrical phenomena in conductive materials.

\subsection{Non-annealed characterization of the $\mathrm{WO}_3$ layers}
 
\subsubsection{Surface morphology: FE-SEM analysis}
\begin{figure}
\begin{center}
\resizebox{0.48\textwidth}{!}{%
\includegraphics{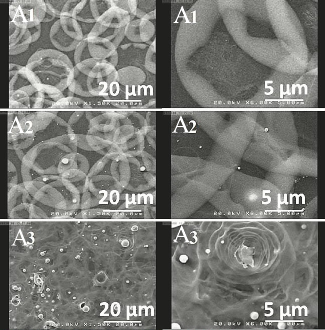}
}
\caption{%
  The FE-SEM images of the prepared $\mathrm{WO}_3$ samples are shown at three different spray rates: R$_1=1$ mL/min, R$_2=3$ mL/min, and R$_3=7$ mL/min, as depicted in Figure \ref{FESEM}.}
\label{FESEM}
\end{center}
\end{figure}

Figure \ref{FESEM} displays FE-SEM images of the grown samples A$_1$, A$_2$, and A$_3$ with their corresponding rates R$_1$, R$_2$, and R$_3$, captured at two scales: 20 $\mu$m and 5 $\mu$m. The FE-SEM images reveal that all samples are coated with numerous flat thin layers, forming spiral strings with rotating clusters known as lattice tungsten filaments. Notably, the presence of tungsten filaments with varying diameters is evident in all layers. The diameter of these filaments decreases with an increase in the deposition rate.
\subsubsection{Structural properties}
\begin{figure}
\begin{center}
\includegraphics[scale=1.45,trim=0 0 20 0, clip]{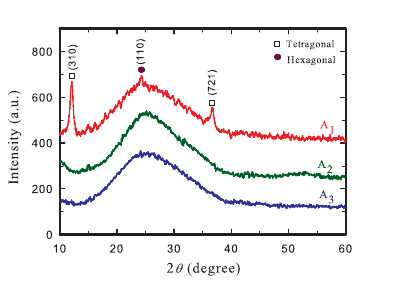}

\caption{%
The XRD spectra of studied $\mathrm{WO}_3$ thin layers. The sample A$_1$ possessing the lowest deposition rate $\mathrm{R}_1$ has been grown with a polycrystalline nature in a mixed tetragonal and hexagonal phases. Other samples were grown in amorphous phase.}
\label{XRD}
\end{center}
\end{figure}

The XRD patterns of the studied layers are shown in Figure \ref{XRD}. Upon inspecting this figure, it is evident that the sample with the lowest deposition rate, R$_1$, exhibits a polycrystalline nature. It consists of both tetragonal and hexagonal phases, with lattice parameters of $a = b = 23.33$ Å and $c = 3.79$ Å for the tetragonal phase, and $a = b = 7.29$ Å and $c = 3.89$ Å for the hexagonal phase. The patterns were indexed using SigmaPlot software, and it was observed that all peaks were in good agreement with the hexagonal structure of WO$_3$ by utilizing the standard data (JCPDS No. 05-0392) \cite{Prabhu2014}. On the other hand, the samples grown at higher deposition rates, R$_2$ and R$_3$, appear to be in the amorphous phase.
Using theoretical analysis associated with the crystalline structure, it is possible to deduce the distance between adjacent planes and the crystallite size in the grown samples. The distance between crystal planes, denoted as $d(hkl)$, can be obtained using Bragg's formula.
\begin{equation}\label{Bragg}
\begin{array}{lcl}
d_{(hkl)}=\dfrac{n\lambda}{2\sin\theta}.
\end{array}
\end{equation}
Accordingly, the crystallite size corresponding to a given Bragg's angle $\theta$ can be found using Scherer's equation.
\begin{equation}\label{Scherer}
\begin{array}{lcl}
D=0.9\frac{\lambda}{\beta\cos\theta},
\end{array}
\end{equation}
where $\lambda$ is the X-ray wavelength and $\beta$ is the full width at half maximum (FWHM) of the given peak. By using Equation (\ref{Scherer}), for the ($3\;1\;0$) peak positioned at the Bragg's diffraction angle of $\theta \approx \ang{12}$, the crystallite size of sample A$_1$ is estimated to be around 16 nm. Additionally, the dislocation density $\delta$ of the sample can be calculated using the formula $\delta = 1/D^2$ \cite{Xin2009} (for more information, see Table \ref{table1}).
\begin{table}[b]
  \caption{Crystallite size, distance between crystal planes, strain, and density of dislocations in sample A$_1$ were analyzed.}
  \begin{tabular}[htbp]{@{}lllll@{}}
    \hline
    Sample & D(nm) & $d_{hkl}$(nm) & $\varepsilon(\times 10^{-3}$) &  $\delta(\times 10^{-3})$(nm)$^{-2}$ \\
    \hline
    A$_1$  & 15.93  & 0.733 & 2.17 & 3.9  \\
    \hline
  \end{tabular}
  \label{table1}
\end{table}
\subsubsection{Electric and thermoelectric properties}

The measurement tool description of the Seebeck effect for electric and thermoelectric properties is given as follows.  
A metal-semiconductor connection can be established to investigate the conductivity of the sample using an electroaccumulation device. This device offers different modes, including Cyclic Voltammetry (CV), Structure Porous Carbon (CPC), Hydrothermal Carbon (CHC), and extra \cite{Zhang2021}. In our study, we specifically employed the CV mode. The device settings involve defining a potential range from E$_1$ to E$_2$ and determining the number of steps for this scanning process as well as the repetitions of this potential cycle. Subsequently, by connecting the electrical terminals of the device to the output terminals of the component and applying voltage within a specific range, we measure the current flowing through the sample.

\begin{figure}
\begin{center}
\includegraphics[scale=1.35,trim=5 5 20 0, clip]{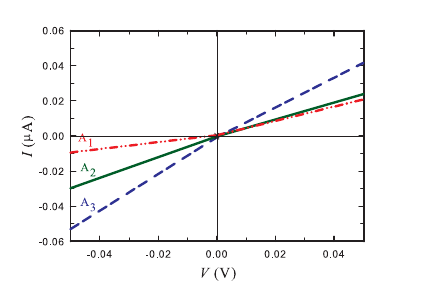}

\caption{%
   The $I-V$ curves of the three studied samples are depicted in Figure \ref{I_V}. The plots exhibit a clear ohmic behavior, indicating that the samples follow Ohm's law.}
\label{I_V}
\end{center}
\end{figure}

\begin{figure}
\begin{center}

\includegraphics[scale=1.3,trim=00 0 5 5, clip]{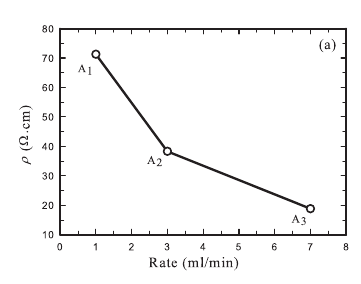}
\includegraphics[scale=1.25,trim=0 00 5 5, clip]{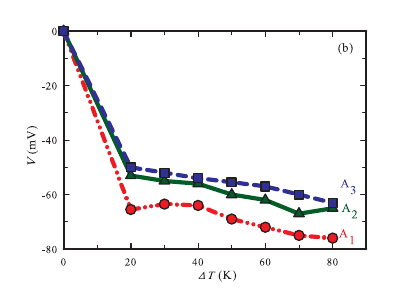}

\caption{(a) The dependence of the electrical resistivity on the spray rate for the samples prepared using a subtractive process. The plot clearly shows a decrease in resistivity as the spray rate increases. This trend can be attributed to an increase in the density of oxygen vacancies within the network layer.
(b) The experimental data of the Seebeck effect in the studied samples. The Seebeck coefficient, which is determined by the majority of carriers, indicates the type of conductivity exhibited by the samples. In this case, all layers exhibit $n$-type conductivity. The temperature difference across the width of the samples was fixed at $\Delta T = \ang{10}$K during the measurements. }
\label{ER_Seebeck}
\end{center}
\end{figure}

Figure \ref{I_V} illustrates the current-voltage curves of the samples under investigation. The plots clearly demonstrate that all samples exhibit an ohmic behavior, and the resistivity of the layers decreases as the spray rate increases. To further elaborate on this phenomenon, we provide the following explanation. The sheet resistance, denoted as $R_{sh}$, can be determined by the ratio of voltage ($V$) to current ($I$), while the resistivity of the layers ($\rho$) is obtained from the relation $\rho = R_s t$, where $t$ represents the thickness of the layer. In our measurements, these quantities were evaluated for a $1\times 1$ cm$^2$ area of the samples.
Figure \ref{ER_Seebeck}(a) displays the resistivity variations of the studied layers. The plot reveals a significant decrease in resistivity from 71.3 $\mathrm{\Omega}\cdot$cm for R$_1$ to approximately 18.8 $\mathrm{\Omega}\cdot$cm for R$_3$, which corresponds to the highest spray rate. This decrease in resistivity can be attributed to an increase in the density of oxygen vacancies within the network layer as the spray rate increases.

To determine the type of conductivity (either $n$-type or $p$-type) exhibited by the samples, we employed the Seebeck effect relation $\mathcal{V}=S\Delta T$ \cite{Ashcroft}, where $S$ is the Seebeck coefficient, $\mathcal{V}$ is the thermoelectric power, and $\Delta T$ is the temperature difference across the width of the sample. The experimental results for the Seebeck effect are presented in Fig. \ref{ER_Seebeck}(b).

Based on these results, it can be concluded that the Seebeck coefficient of the samples is negative, indicating $n$-type conductivity. Previous studies have reported an inverse proportionality between the Seebeck coefficient and the conductivity of the layer \cite{Owen1978}. This finding is consistent with our obtained values from the $I-V$ data for the sheet resistance of the samples.

\subsubsection{Optical properties}
\begin{figure}
\begin{center}
\includegraphics[scale=1.45,trim=0 0 5 0, clip]{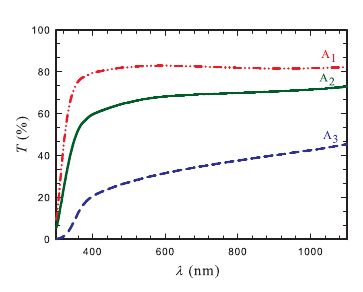}

\caption{%
 The UV--Vis. transmittance spectra of the $\mathrm{WO}_3$ films prepared under different spray rates (1 mL/min $\leq R \leq$ 7 mL/min) are shown in Figure \ref{UV_Vis}. The spectra illustrate the variation in transmittance across the UV and visible regions for the samples.}
\label{UV_Vis}
\end{center}
\end{figure}

\begin{figure}
\begin{center}
\includegraphics[scale=1.35,trim=0 0 0 0, clip]{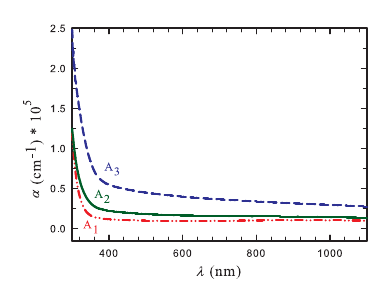}

\caption{%
  The absorption coefficient as a function of wavelength $\lambda$. In the ultraviolet region ($\lambda <$ 400 nm), all samples exhibit high absorption coefficients, approximately on the order of $10^4$ cm$^{-1}$.}
\label{Abs_Coef}
\end{center}
\end{figure}

\begin{figure}
\begin{center}

\includegraphics[scale=1.3,trim=00 00 20 10, clip]{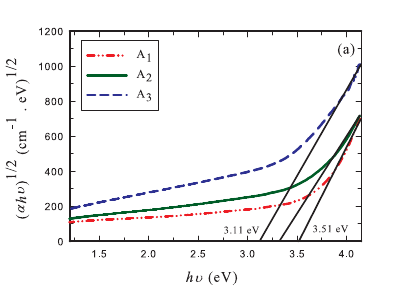}
\includegraphics[scale=2.68,trim=0 00 0 5, clip]{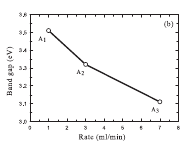}

\caption{%
  The details of the band gap determination: (a) indirect band gaps (b) The final calculated results and variations of $E_g$.}
\label{E_g}
\end{center}
\end{figure}

Figure \ref{UV_Vis} displays the UV-Vis. transmittance spectra of the non-annealed samples. In the visible region (around $t \approx$ 550 nm), the transmittance values are approximately 80$\%$ for A$_1$, 70$\%$ for A$_2$, and 30$\%$ for A$_3$. These variations in transmission are consistent with the improvement in the crystallinity of the layers and their thicknesses (approximately $t \approx 200 \pm 20$ nm, $235 \pm 20$ nm, and $290 \pm 20$ nm for the corresponding rates R$_1$, R$_2$, and R$_3$, respectively).
The observed changes in transmittance can be attributed to the reduction in photon scattering due to the better crystalline arrangement of the layers. As studied by M. Regragui {\it et al.} \cite{Regragui2000}, the transmittance behavior is influenced by the annihilation of photon scattering, which is facilitated by the improved crystallinity in the layers. Consequently, we observe an enhancement in light transmission through the layers.
Figure \ref{Abs_Coef} depicts the spectra of the absorption coefficient of the samples as a function of wavelength $\lambda$, obtained using Lambert's equation.
\begin{equation}
\begin{array}{lcl}
\alpha=-\frac{\ln T}{t}
\end{array}
\end{equation}
It is evident from this figure that all three samples exhibit relatively high absorption coefficients, approximately on the order of $10^4$ cm$^{-1}$, in the ultraviolet region ($\lambda <$ 400 nm). Consequently, sample A$_1$ grown at rate R$_1$ displays a sharp absorption edge at 320 nm. As the spray rate increases, the absorption edge shifts towards longer wavelengths. By utilizing this information, we can determine the optical band gap $E_g$ of the investigated films using the following formula: \cite{Bathe2007} 
\begin{equation}
\begin{array}{lcl}
(\alpha h\nu)^m=A(h\nu - E_g)
\end{array},
\end{equation}
Here, $A$ represents a constant, $h\nu$ denotes the incident photon energy, and $m$ depends on the nature of the band transition ($m=2$ for direct transitions and $m=1/2$ for indirect ones). The determination of both direct and indirect band gaps involves extrapolating the linear portion of the curve to the energy axis, specifically $(\alpha h\nu)^m=0$. The analysis for the indirect band gaps is illustrated in detail in Fig. \ref{E_g} (a), and the final results are presented in Fig. \ref{E_g} (b).
We observed that $E_g$ increases from 3.11 to 3.51 eV as the spray rate decreases. These variations can be explained by the width of the distribution of defect states (mainly oxygen vacancies) near the conduction or valence band edges.

\subsection{Annealed $\mathrm{WO}_3$ layers characterization}
\subsubsection{Structural properties}
$\mathrm{WO}_3$ thin films were deposited on glass substrates using the spray pyrolysis technique at three different rates: R$_1=1$ mL/min, R$_2=3$ mL/min, and R$_3=7$ mL/min. Subsequently, the films were annealed for 2 hours at $\ang{500}$C in air. XRD structural analysis revealed that the sample $\mathrm{A}_2$ grown at a rate of R$_2=3$ mL/min exhibited an amorphous nature, while samples $\mathrm{A}_1$ (R$_1=1$ mL/min) and $\mathrm{A}_3$ (R$_3=7$ mL/min) were found to be polycrystalline.
\begin{figure*}
\begin{center}
\includegraphics[scale=0.95,trim=0 0 5 10, clip]{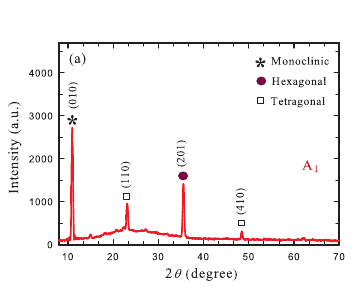}
\includegraphics[scale=0.9,trim=0 0 5 10, clip]{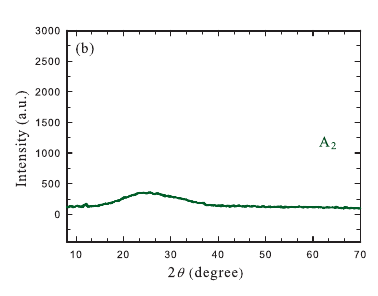}
\includegraphics[scale=0.88,trim=0 0 5 0, clip]{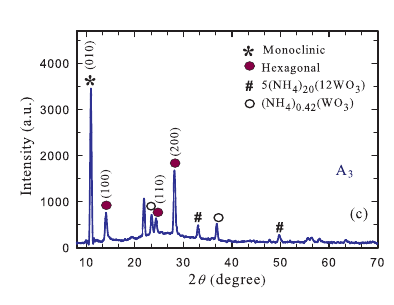}

\caption{%
XRD diagrams of $\mathrm{WO}_3$ thin layers prepared at $T_R = \ang{400}$C and subsequently annealed at $T_A = \ang{500}$C are shown in (a) A$_1$, (b) A$_2$, and (c) A$_3$. After annealing, A$_1$ and A$_3$ exhibit a polycrystalline structure, while A$_2$ remains in the amorphous phase.
}
\label{XRD_Ann}
\end{center}
\end{figure*}
Figure \ref{XRD_Ann} displays the crystalline structure of thin films after annealing. It is noteworthy that before annealing  two samples with rates $\mathrm{R}_2$ and $\mathrm{R}_3$ had an amorphous nature, while the sample $\mathrm{A}_1$ with rate $\mathrm{R}_1$  had polycrystalline monoclinic structure. As shown in Fig. \ref{XRD_Ann} we see that after annealing samples $\mathrm{A}_1$ and $\mathrm{A}_3$ have polycrystalline structure, whereas the sample $\mathrm{A}_2$ remains in the amorphous phase. 

Interestingly, the sample $A_1$ is  grown in three phases: monoclinic phase with lattice parameters $a=14.49 \si{\angstrom}$, $b=14.50 \si{\angstrom}$, $c=10.92 \si{\angstrom}$, hexagonal phase with lattice parameters $a=b=7.28 \si{\angstrom}$, $c=3.88 \si{\angstrom}$ and tetragonal phase with lattice parameters $a=b=7.56 \si{\angstrom}$, $c=3.73 \si{\angstrom}$. 
Moreover, $A_3$ sample is grown in the two aforedescribed phases: monoclinic with lattice parameters $a=14.49 \si{\angstrom}$, $b=14.50 \si{\angstrom}$, $c=10.92 \si{\angstrom}$ and hexagonal with lattice parameters $a=b=7.29 \si{\angstrom}$, $c=3.89 \si{\angstrom}$ together with two new phases so-called ($5(\mathrm{NH}_4)_{20}) .12\mathrm{WO}_3$) and ($(\mathrm{NH}_4)_{0.42}\mathrm{WO}_3$).
The preferential directions for both polycrystalline samples are grown along ($0\;1\;0$) direction positioned at the Bragg angle of
 $\theta\approx \ang{12}$ (JCPDS 05-0392). Other researchers have reported the same results for example in Refs.\cite{Bertus2012,Dabbous2009}. 
 
In order to further investigation of  the obtained data, one can calculate crystal size, distance between network planes, strain and density of dislocation $\delta$ by implementing equations (\ref{Bragg}) and (\ref{Scherer}).
\begin{table}[b]
  \caption{Crystallite size, distance between the crystal planes, strain and the density of dislocation of planes in the samples A$_1$ and A$_3$ after annealing.}
  \begin{tabular}[htbp]{@{}lllll@{}}
    \hline
    Sample & D(nm) & $d_{hkl}$(nm) & $\varepsilon(\times 10^{-3}$) &  $\delta(\times 10^{-3})$(nm)$^{-2}$ \\
    \hline
    A$_1$  & 30.66  & 0.804 & 1.14 & 1.06  \\
    A$_3$  & 39.42  & 0.814 & 0.87 & 0.64  \\
    \hline
  \end{tabular}
  \label{table2}
\end{table}  
Table \ref{table2} indicates that after annealing, the crystal size increases with an increase in the spray deposition rate, except for sample A$_2$ which remains in an amorphous phase. This phenomenon is consistent with the results reported in previous works \cite{Abe2008, Sun2009, Chacko2006}. It is worth noting that the distance between network planes is well-matched across the samples. 
It is interesting to mention that $\text{WO}_3$ exhibits unconventional behavior in the laboratory, and its structural properties studied by spray pyrolysis are unpredictable, resulting in either amorphous or crystalline structures.
From the structural analysis results, we can calculate the strain $\varepsilon = \beta \cos(\theta)/4$ and the dislocation density $\delta = 1/D^2$ of the crystal plates \cite{Sawaby2010}. The dislocation density represents the length of dislocation lines per unit volume of the crystal.
Moreover, as the deposition rate increases, both the strain and the density of dislocations remarkably decrease, indicating improved crystallinity of the layers after annealing. This observation is consistent with the density of the ($0\;1\;0$) peak \cite{Abe2008, Sun2009}.

\subsubsection{Electrical and Thermoelectric Properties}
\begin{figure}
\begin{center}
\resizebox{0.5\textwidth}{!}{%
\includegraphics{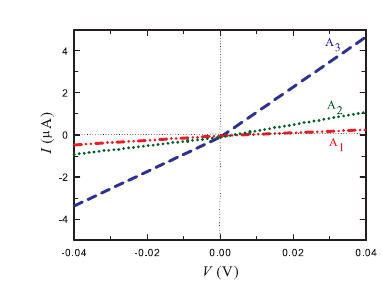}
}
\caption{%
 $I-V$ characteristics of the studied samples after annealing. The plots exhibit clear Ohmic behavior, indicating a linear relationship between current and voltage for the samples.}
\label{I_V_Ann}
\end{center}
\end{figure}

\begin{figure}
\begin{center}

\includegraphics[scale=1.3,trim=00 5 5 6, clip]{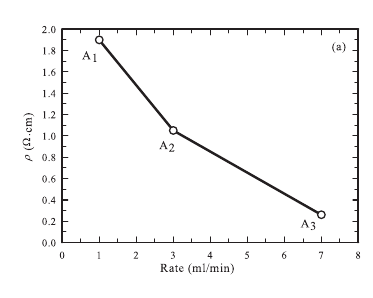}
\includegraphics[scale=1.5,trim=00 5 5 5, clip]{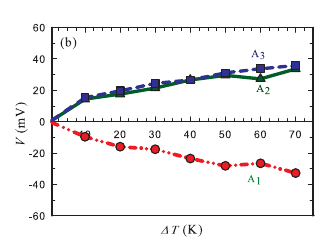}

\caption{%
(a) Electrical resistivity of samples after annealing for 2 hours at $\ang{500}$C.
(b) Experimental data of the Seebeck effect. The temperature difference $\Delta T = \ang{10}$C was maintained constant throughout the experiment.}
\label{R_Seebeck_Ann}
\end{center}
\end{figure}

Figure \ref{I_V_Ann} displays the $I-V$ curve of the studied samples after annealing. The sheet resistance $R_{sh}$ of the layers decreases from 84 $\mathrm{k\Omega}$ to 8.3 $\mathrm{k\Omega}$, indicating an improved electrical conductivity. Additionally, we observe a decrease in resistivity $\rho$ from 1.9 $\mathrm{\Omega}\cdot\mathrm{cm}$ to 0.26 $\mathrm{\Omega}\cdot\mathrm{cm}$, as shown in Figure \ref{R_Seebeck_Ann} (a).
The decrease in $R_{sh}$ can be attributed to two reasons: firstly, the increase in carrier mobility, which depends on the dislocation density, and secondly, an increase in carrier density. The latter corresponds to an increase in acceptor concentration (tungsten vacancies/holes) relative to oxygen vacancies with an increase in both the spray deposition rate and annealing. \cite{Regragui2000,Dickens1968,Lightsey1973,Crandall1977,Owen1978}. 
The experimental results for the Seebeck effect of the samples after annealing are presented in Fig. \ref{R_Seebeck_Ann} (b). The data clearly indicate that the majority of carriers in sample $A_1$ are electrons, indicating its $n$-type conductivity. On the other hand, samples $A_2$ and $A_3$ exhibit majority carriers as holes, indicating their $p$-type conductivity.

The observed conductivity types are a result of the deposition and annealing processes being performed in ambient air. During these processes, the surface of the $\mathrm{WO}_3$ films can absorb oxygen, leading to the creation of tungsten vacancies with positive holes. The density of donor-like levels, particularly oxygen vacancies, decreases while the density of acceptor-like levels, specifically tungsten vacancies, increases. This phenomenon has been previously reported in the literature \cite{Patila2000, Gilleta2004}.

It is known that the $n$-type conductivity of $\mathrm{WO}_3$ is primarily attributed to the presence of oxygen vacancies, which are associated with non-stoichiometry in the $\mathrm{WO}_3$ lattice. The balance between the dopant concentration and oxygen deficit influences the $n$-type conductivity of the $\mathrm{WO}_3$ layers. In our case, the deposition and annealing in an oxidizing environment (air) result in the reduction of oxygen vacancies during the annealing process. This reduction contributes to the generation of positive holes, leading to a switch from $n$-type conductivity to $p$-type conductivity. 

Furthermore, the highly oxidizing atmosphere can cause oxygen atoms to adsorb onto the surface of the $\mathrm{WO}_3$ film, creating tungsten vacancies with six positive holes. This phenomenon is responsible for the $p$-type conductivity typically observed in $\mathrm{WO}_3$ thin films \cite{Bertus2012}.

\subsubsection{Optical properties }
\begin{figure}
\begin{center}
\resizebox{0.5\textwidth}{!}{%
\includegraphics{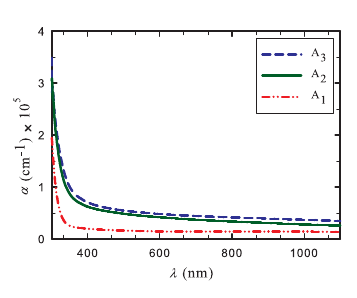}
}
\caption{%
The variation of absorption versus wavelength. }
\label{Var_Wave}
\end{center}
\end{figure}

The optical properties of the annealed $\mathrm{WO}_3$ thin films were extensively studied. 
After annealing, the absorption coefficient ($\alpha$) of the samples increases compared to the non-annealed stage. This indicates that the studied samples exhibit high absorbance properties ($\alpha \approx 10^5$ cm$^{-1}$) in the ultraviolet wavelength region ($\lambda \leq 400$ nm). Figure \ref{Var_Wave} illustrates the absorption as a function of wavelength ($\lambda$) for the annealed samples.

\begin{figure}
\begin{center}

\includegraphics[scale=1.25,trim=00 5 5 5, clip]{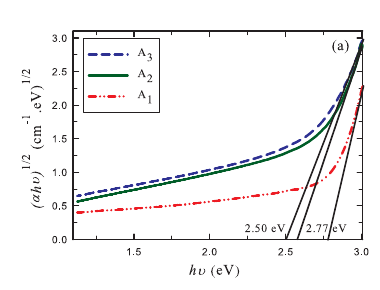}
\includegraphics[scale=1.29,trim=00 5 5 5, clip]{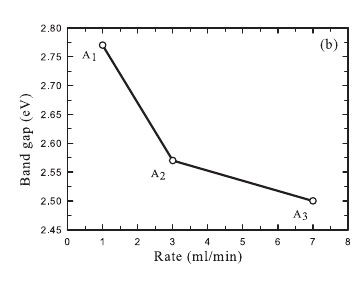}

\caption{%
  (a)Indirect band gaps ($E_g$) calculated from the absorption coefficient of the layers. The spray deposition rate influences the indirect optical band gap of the layers.
(b) Comparison of the indirect band gaps ($E_g$) for different spray deposition rates. The optical band gap of these layers decreases with an increase in the spray deposition rate.}
\label{Fig13_Bnd_Gap}
\end{center}
\end{figure}

Using the absorption coefficient, we can straightforwardly determine the indirect optical band gap of the layers, as depicted in Figs. \ref{Fig13_Bnd_Gap} (a) and (b). From these figures, it can be observed that the band gap of the considered layers decreases from 2.77 eV in A$_1$ to 2.50 eV in A$_3$ as the spray deposition rate increases.

The quantum confinement effect (QCE) describes electrons in terms of energy levels, potential wells, electron energy band gaps, etc. It is observed when the size of the particle is too small to be comparable to the electron wavelength. WO$_3$ exhibits attractive properties due to the QCE caused by its small particle size (usually $<$10 nm) \cite{Wang2017}.

Furthermore, our investigations reveal that the reduction in band gap mentioned earlier is related to an increase in crystal size and may be attributed to the attenuation of the quantum confinement effect \cite{Owen1978}.

\section{Conclusion}
In this study, we investigated the influence of the deposition rate and annealing on the physical properties of $\mathrm{WO}_3$ thin films prepared by the spray pyrolysis method. The FE-SEM images revealed the formation of circular strings with varying diameters, which decreased as the deposition rate increased. The XRD analysis indicated that sample A$_1$ exhibited a polycrystalline structure with mixed hexagonal-tetragonal phases. 

Analysis of the optical data showed a decrease in the indirect band gap of the layers from 3.51 eV to 3.11 eV, attributed to changes in the width of the band tails of oxygen vacancies defect states near the band edges. 

Regarding the Seebeck effect, our experimental results indicated that sample A$_1$ exhibited $n$-type conductivity, with electrons being the majority carriers, while samples A$_2$ and A$_3$ exhibited $p$-type conductivity, with holes being the majority carriers. After annealing, samples A$_1$ and A$_3$ exhibited crystallization in different structures. Additionally, the annealed films displayed a relatively high absorption coefficient of approximately $10^5 \; \mathrm{cm}^{-1}$ and an indirect transition gap within the range of 2.50-2.77 eV for samples A$_1$ and A$_3$. Furthermore, we observed that these materials exhibited $n$-type characteristics after annealing.

Based on our analysis, we have demonstrated the potential of the Seebeck effect as a useful experimental tool for determining the conductivity type of $\mathrm{WO}_3$ under different conditions. As part of our future plans, we are currently conducting electrical measurements to optimize the resistivity of these thin films.

\section*{Acknowledgements}

Hamid Arian Zad acknowledges the receipt of the grant from the Abdus Salam International Centre for Theoretical Physics (ICTP), Trieste, Italy, and the CSMES RA in the frame of the research project No. SCS 19IT-008.
 H.A.Z also acknowledges for the financial support of the National Scholarship Programme of the Slovak Republic (N{\v S}P). 

\section*{Funding}
This study was funded by ICTP and SAIA.

\section*{Conflicts of interest}
The authors have no conflicts of interest to declare that are relevant to the content of this article.

\section*{Author contribution statement}
All authors contributed equally to this work.

\bibliography{}

\end{document}